\documentclass[useAMS,usenatbib]{mn2e}
\usepackage[dvips]{graphicx}
\usepackage{color}

\title[Point-source parameters for scanning surveys]{Optimizing point-source parameters for scanning satellite surveys}
\author[F. van Leeuwen and A. N. Morgan and D. L. Harrison]{F. van Leeuwen$^{1}$\thanks{E-mail: fvl@ast.cam.ac.uk (FVL)} and A. N. Morgan$^{1,2}$\thanks{E-mail: amorgan@astro.berkeley.edu (ANM)} and D. L. Harrison$^{1}$\thanks{E-mail: dlh@ast.cam.ac.uk (DLH)}\\
$^{1}$Institute of Astronomy, Madingley Road, Cambridge, CB3 0HA, UK\\
$^{2}$Department of Astronomy, 601 Campbell Hall, University of California at Berkeley, Berkeley, CA, 94720, USA}
\begin{document}

\date{Accepted . Received ; in original form }

\pagerange{\pageref{firstpage}--\pageref{lastpage}} \pubyear{2009}

\maketitle

\label{firstpage}

\begin{abstract}
We describe a method for deriving the position and flux of point and compact sources observed by a scanning survey mission. Results from data simulated to test our method are presented, which demonstrate that at least a 10-fold improvement is achievable over that of extracting the image parameters, position and flux, from the equivalent data in the form of pixel maps. Our method achieves this improvement by analysing the original scan data and performing a combined, iterative solution for the image parameters. This approach allows for a full and detailed account of the point-spread function, or beam profile, of the instrument. Additionally, the positional information from different frequency channels may be combined to provide the flux-detection accuracy at each frequency for the same sky position. Ultimately, a final check and correction of the geometric calibration of the instrument may also be included. The {\it Planck} mission was used as the basis for our simulations, but our method will be beneficial for most scanning satellite missions, especially those with non-circularly symmetric point-spread functions.

\end{abstract}


\section{Introduction}

Survey missions designed primarily for observing the continuum radiation at one or more frequencies (for example WMAP \citep{bennett97}, Planck \citep{tauber04}), generally produce as a by-product a point- or compact-source catalogue \citep[e.g.\ ][]{lopez07,vielva03}, containing positional and flux information on the detected sources. Generally these data have been, or are planned to be, extracted from the pixel maps on which the survey data are projected and collected. The process involves a point-source detection on those maps, followed by an estimate of the position and flux from the pixels. There are two intrinsic problems here. Firstly, each pixel will represent a different collection of scan directions. Secondly, the digitisation on the pixel map will lead to signal distortion, as centres and sizes of map-pixels will never coincide accurately with the original size and shape of the measured samples. When the point-spread function (PSF) of the instrument is circular symmetric, these problems may be partly overcome. However, any asymmetry in the PSF will lead to a distortion of the accumulated image that is difficult, if not impossible, to fully incorporate when deriving from the pixel maps the all-important image parameters: the position and flux of the source with their formal errors. It was also considered that pixelisation of the data could have a significant effect on the derived source parameters and their standard errors.

Providing accurate and fully internally consistent positional and flux information for  point sources detected in a survey mission is crucially important for any subsequent use of those data. Cross identification followed by flux comparisons for different wavelengths is only the most obvious application affected. Studies of the different spectral shapes among the detected sources will rely on the compatibility and accuracy of data obtained in different wavelengths to allow for reliable classification of (newly discovered) objects.

We started the current study with the following hypothesis: \textit{The transit data used to build the point-source image contain together all information (image parameters) on the source, convolved for each transit with a PSF that is known or possible to reconstruct. Extracting the image parameters directly from the accumulated transit data should therefore make it possible to account directly for the applicable PSF for each transit.} The question we then asked ourselves was the following: \textit{When extracting the image parameters directly from the transit data, will we see a significant improvement in the accuracy and the statistics of the resulting image parameters, compared to extraction from the accumulated data on a pixel map?} To test this, one of us (ANM) prepared a software package with which the image parameter extraction methods for pixel maps and for combined transit data could be tested and compared, while DLH prepared simulated data to be input to the tests, following methods we used in an earlier study \citep{dlh05}. 

The technique for extracting image parameters as presented here, can be seen as a simplified version of the solution that has been applied to the Hipparcos data reductions \citep{fvl07}. Similar to those reductions, it can in principle also include a geometric calibration, which defines the relations between detector pointings at the sky and the satellite attitude. Similarly, as part of an iterative procedure, these methods can be used to prepare data as input for the calibration of the PSF. These kinds of dependencies are common for all-sky survey satellites, for which the data reduction is best treated as ``self calibrating", to provide well defined internally rigid data systems. Such systems can then be calibrated as a whole to externally defined reference values. However, these calibration aspects are not treated in this paper, which concentrates on the extraction of the image parameters only.

This paper has been organized as follows. Section~\ref{sec:technique} presents the mathematical background for the methods used, followed by a description of the data simulations in Section~\ref{sec:datasim}. The processing methods for the scan data to image parameters is presented in Section~\ref{sec:scandata}, and for the pixel maps in Section~\ref{sec:pixmapdata}. The results of the application of these methods and a discussion of the results are presented in Section~\ref{sec:comparisons} and Section~\ref{sec:discussion}. 

We use ecliptic coordinates ($\lambda,\beta$) throughout the paper, because full-sky scanning strategies for satellites are generally related to the ecliptic-plane position of the Sun, and tend to have symmetries with respect to the ecliptic plane.

\section{The mathematical model}
\label{sec:technique}

\subsection{The image response function}
The methods we describe here are applicable to sources for which the apparent diameter on the sky is small with respect to the PSF of the detector, but can in principle be extended to include sources for which the shape can, in first approximation, be represented by a known circular symmetric distribution. In both cases an image response function (IRF), $R$, can be created that predicts the image shape as the response of the instrument to an observation of a source. 

The IRF is described as an arbitrary function of the along- and across-scan coordinates, $\Delta\upsilon\equiv\upsilon_t-\upsilon_s$ and $\tau_t$. Here $\upsilon_s$ refers to the scan phase of the detector. The subscript $t$ refers to the coordinates of a target source, measured relative to the scan as described by the detector:
\begin{equation}
R = R(\Delta\upsilon, \tau).
\end{equation}  
The IRF incorporates the integration over fixed-length samples along-scan, and is thus, for a point source, the actual instrumental PSF convolved with a single-sample width block function. For the current application, the IRF is limited to a suitable radius, referred to as the image perimeter. The IRF is normalized in a relatively arbitrary way, for example by defining the peak to be at a level equal to 1.0. Similarly, the position of the peak can be defined as the centre of the IRF. Whatever way is chosen, it defines the observational response scale for the flux measurements, and the reference frame for the geometric calibration of the detector. The image response for an observed point source is simply given by the IRF times the intensity or flux of the source. A typical transit for a source will consist of the responses for a sequence of equally spaced samples as obtained for a set of scan phases $\upsilon_s$, roughly centred around the along-scan position of the source, $\upsilon_t$. This signal then includes the response from the source at the across-scan position of $\tau_t$. The scan circle itself is defined, for example, by the path described by the maximum possible response direction for a detector, and is usually close, but not equal, to a great circle. 

\subsection{Projection effects}
\label{sec:projeffects}
The geometric calibration of an instrument such as the Planck satellite has been described in detail in \citet{dlh05} and \citet{dlh06}, where it is shown that an accuracy of about 0.01 times the FWHM (full-width half maximum) of the IRF can realistically be obtained. Thus, in the following discussion, uncertainties in the geometric calibration will in first instance be ignored, though we will come back to these very briefly in the final section of this paper. Similarly, the satellite attitude (in this case the position of the spin axis and the scan phase) can quite easily be determined to a relatively high accuracy, as this will be based on a star mapper and a high accuracy star catalogue. Thus, we will assume errors on the knowledge of the position of the detector as a function of time to be negligible with respect to other uncertainties, the most important of which will be the photon noise (or equivalent) on the original measurements.

At each stage of the analysis we assume to have available an estimated position and flux of the source we are investigating. In first instance this may simply come from the original identification of the source on a pixel map. The discrepancy between the actual and estimated positions and fluxes of the source will reflect in systematic differences between the predicted and observed responses. These differences together contain the information required for correcting the original to the best estimates of the source's position with respect to the scan ($\upsilon_t,\tau_t$) and flux.

The corrections to the estimated sky coordinates $\tilde{\lambda}_t,\tilde{\beta}_t$ of the source are reflected in corrections to the $\upsilon_t$ and $\tau_t$ values for each scan, in a way that depends on the local orientation of the scan circle. The combination of scan circles with different local orientations allows for the reconstruction of the actual positional corrections. The local orientation of the scan circle is defined by the instantaneous position of the spin axis for the scan ($\lambda_p,\beta_p$), the opening-angle of the scan (the angle $\alpha$ between the detector and the spin axis) and the position of the source ($\lambda_t,\beta_t$). The local inclination of the scan $\psi$ is now defined by:
\begin{eqnarray}
\sin\psi &=& \frac{-\sin(\lambda_t-\lambda_p)\cos\beta_p}{\sin\alpha}, \nonumber \\
\cos\psi &=& \frac{\sin\beta_p - \sin\beta_t\cos\alpha}{\cos\beta_t\sin\alpha},
\label{equ:defpsi}
\end{eqnarray}
with which the relations between $\Delta\upsilon$, $\tau$ and the corrections to the source-coordinates are expressed as:
\begin{eqnarray}
\mathrm{d}\upsilon_t &=& \phantom{-}\mathrm{d}\lambda_t\cos\beta_t\cos\psi +\mathrm{d}\beta_t\sin\psi, \nonumber \\
\mathrm{d}\tau_t &=& -\mathrm{d}\lambda_t\cos\beta_t\sin\psi +\mathrm{d}\beta_t\cos\psi.
\label{equ:transf}
\end{eqnarray}

\subsection{The observations}
In the following we examine the data collected for a point source. We assume that we have an accurate predicted image shape, which is equivalent to the IRF for a flux $I$:
\begin{equation}
R_P = I\cdot R(\upsilon-\upsilon_s,\tau).
\end{equation}
This predicted image shape describes for each transit of the source the expected response as a function of the along- and across-scan positions of the source with respect to the scan circle. The signal is sampled at a series of discrete values $j$ of $\upsilon_s$, to provide observations:
\begin{equation}
o_j = I\cdot R(\upsilon_t-\upsilon_{s,j},\tau_t) + \epsilon_j,
\end{equation}  
where $\epsilon_j$ will generally be a function of the signal strength itself (such as Poisson noise).

The predicted position, together with the attitude details of the scans, defines for each scan predicted scan position of the source, $\tilde{\upsilon}_t$ and $\tilde{\tau}_t$. Together with the predicted flux, this defines predicted observations for each scan, which can be compared with the observed counts:
\begin{equation}
\tilde{o}_j = \tilde{I}\cdot R(\tilde{\upsilon}_t-\upsilon_{s,j},\tilde{\tau}_t),
\end{equation}  

\subsection{Correcting estimated values}
\label{sec:correst}
The differences between the predicted and observed instrument responses can be expressed as a first-order approximation to corrections for $\tilde{\upsilon}_t$, $\tilde{\tau}_t$ and $\tilde{I}$:
\begin{eqnarray}
o_j - \tilde{o}_j &=& \mathrm{d}I\cdot R(\tilde{\upsilon}_t-\upsilon_{s,j},\tilde{\tau}_t) \nonumber \\
&+& \tilde{I}\cdot\frac{\partial R}{\partial\upsilon_t}\:\mathrm{d}\upsilon_t  
+ \tilde{I}\cdot\frac{\partial R}{\partial\tau_t}\:\mathrm{d}\tau_t + \epsilon_j,
\label{equ:observ}
\end{eqnarray}
where $\upsilon_t = \tilde{\upsilon}_t+\mathrm{d}\upsilon_t$, and similarly for the other variables. By combining Eq.~\ref{equ:transf} and Eq.~\ref{equ:observ}, the corrections to the assumed position and flux of the source, as derived from the observations, are described as a set of linear equations, which can be solved by least squares:
\begin{eqnarray}
o_j - \tilde{o}_j &=& \mathrm{d}I\cdot R(\tilde{\upsilon}_t-\upsilon_{s,j},\tilde{\tau}_t) \nonumber \\
&+& \tilde{I}\biggl[\frac{\partial R}{\partial\upsilon_t}\cos\beta_t\cos\psi - \frac{\partial R}{\partial\tau_t}\cos\beta_t\sin\psi\biggr]\mathrm{d}\lambda_t \nonumber \\
&+& \tilde{I}\biggl[\frac{\partial R}{\partial\upsilon_t}\sin\psi + \frac{\partial R}{\partial\tau_t}\cos\psi\biggr]\mathrm{d}\beta_t + \epsilon_j.
\label{equ:obs_full}
\end{eqnarray}
Potential weakness or ambiguity in the solution is resolved by the different orientations $\psi$ of the scan circles used, and the different values of the across-scan positions of the source as applicable to the different scans. Some iteration is still required, as corrections to the three estimated image parameters also affect the coefficients in Eq.~\ref{equ:obs_full}. 

\subsection{Extension to multiple frequencies}

The extension from a single to multiple frequency maps is simple, as long as the detectors for the different frequencies have been accurately calibrated with respect to each other. In that case, Eq.~\ref{equ:obs_full} is extended with a correction $\mathrm{d}I$ for each frequency (or possibly for each detector if there are suspected to be response differences between different detectors at the same frequency). The IRF has to be adapted for each detector, but the corrections $\mathrm{d}\lambda_t$ and $\mathrm{d}\beta_t$ are the same for all frequencies. Extension to multiple frequencies could bring improvement in positional accuracy, and so indirectly also to flux estimates.

\subsection{Solving for a background gradient}

A linear 2-dimensional gradient in the background can be resolved through a minor extension of Eq.~\ref{equ:obs_full}. The reason for solving for a possible linear background gradient is its effect on the estimated position of the source, which will get biased by an unsolved background gradient. The background gradient is expressed in offsets for the sky-coordinates of the source:
\begin{equation}
B = B_0 + \frac{\partial B}{\partial (\lambda\cos\beta)}\mathrm{d}\lambda\cos\beta +
\frac{\partial B}{\partial\beta}\mathrm{d}\beta.
\label{equ:background}
\end{equation}
The quantities $\mathrm{d}\lambda\cos\beta$ and $\mathrm{d}\beta$ are directly related to the position of the detector relative to the source through the orientation angle $\psi$ of the scan as defined in Eq.~\ref{equ:defpsi}: 
\begin{eqnarray}
\mathrm{d}\lambda\cos\beta &=& \cos\psi(\upsilon_{s,j}-\upsilon_t) - \sin\psi(\tau_t) \nonumber \\
\mathrm{d}\beta &=& \sin\psi(\upsilon_{s,j}-\upsilon_t) + \cos\psi(\tau_t)
\label{equ:backgrtransf}
\end{eqnarray}
Substituting Eq.~\ref{equ:backgrtransf} into Eq.~\ref{equ:background} describes the background contribution as a function of the source observation coordinates. As such it can be added to Eq.~\ref{equ:obs_full} as an additional set of unknowns to be solved. Some care is required here though, as adding more degrees of freedom has a potential for de-stabilizing the solution, in particular when relatively few independent observations are available. Independent here refers primarily to the distribution of scan direction. The standard errors on the background coefficients as derived from the least squares solution, as well as the standard deviation of that solution, can be used as criteria to decide whether adding a background gradient as additional parameters to the solution is at all beneficial. 

\section{The data simulations}
\label{sec:datasim}
\subsection{Scanning law}
\label{sec:scanlaw}
Full-sky scanning missions with severe constraints on instrument temperature control, such as Hipparcos \citep{fvl07}, Planck, Gaia \citep{lindeg08}, have scanning laws implemented that maintain a fixed angle between the spin axis of the satellite and the direction to the Sun. This angle is referred to as the solar aspect angle $\xi$. The full-sky scanning is obtained by two motions: a precession of the spin axis around the (nominal) direction of the Sun, and a rotational motion of the instrument around the spin axis. This type of scanning strategy has been referred to as ``cycloidal". The limitation of such a scan is the coverage close to the ecliptic plane, for which the spread in inclination angles for the scan coverage is limited by $\xi$. If the inclination of a scan circle with respect to the ecliptic plane is given by $\varphi$ (with $\varphi=\pm\pi/2$ for a scan circle crossing the ecliptic plane at right angles) then
\begin{equation}
|\cos\varphi|\le\frac{\sin\xi}{\sin\alpha},
\end{equation}
where $\alpha$ is the opening-angle of the detector as defined above in Section~\ref{sec:projeffects}. Thus, for a small solar aspect angle, the inclinations of the scan circles remain all close to perpendicular to the ecliptic plane. This is a relatively unfavourable situation for point- and compact-source analysis as the easier to extract positional information is in the along-scan direction. This is therefore a good configuration to see how the current analysis will perform under difficult conditions. For missions such as Hipparcos and Gaia, which aim specifically at the astrometry of point sources, the Solar-aspect angle has been maximized, at values of 43 and 45 degrees respectively \citep{fvl97,lindeg08}, while for Planck it has been chosen at a relatively small value of 7.5 degrees \citep{dupac05}. The Planck mission has therefore been chosen as an example case for our experiments, as it provides a wide range of conditions depending on the ecliptic latitude of the source. The precession period for the spin axis of the Planck satellite is scheduled to be 6 months. 

The Planck scanning \citep{dupac05} is described not as a continuous function, as for Hipparcos and Gaia, but step-wise. This means that the spin axis remains at a ``fixed" position over a period of, for example, 1~hour, during which the satellite describes 60 full circles. The nominal scan velocity is 6 degrees per second, and every part of the scan is examined by 60 successive scans. The accumulated data from the scan circles are referred to as a ``ring". Due to the time required for repositioning the spin axis, there are effectively only about 55 useable scans per pointing. Scans in successive pointings will have a maximum separation of about $2.5\pm0.65$~arcmin. The mean value is the mean motion of the Sun on the ecliptic, the range is the result of precession of the spin axis around the Sun, as a result of which the scan density will be higher when the spin axis moves in opposite direction of the Sun and vice versa. For our experiments we have used one year of ``mission data'', providing a generally homogeneous scan coverage of the sky. 

\subsection{The detectors}
For testing the methods described in Section~\ref{sec:technique} we used a ``worst-case" scenario by selecting the 30~GHz detectors, which have the largest beam size (FWHM$=33$~arcmin) and the largest predicted beam ellipticity (1.4). For this detector we adapted an opening-angle of 85 degrees (this is for testing purposes only and doesn't represent the actual detector position in the Planck focal plane). At the sampling rate of 32.5~Hz, the samples are spaced 5.4~arcmin apart, and a point-source image is effectively covered by 6 to 8 samples.  

The IRF (see section~\ref{sec:technique}) has been represented by a 2-dimensional Gaussian distribution:
\begin{equation}
R(\upsilon,\tau) = \exp\biggl[-\frac{\Delta\upsilon^2}{2\sigma^2_\upsilon}-\frac{\tau^2}{2\sigma^2_\tau}\biggr],
\label{equ:2dgauss}
\end{equation}
where $\sigma=0.5/\sqrt{2\ln 2}\approx 0.425$~FWHM. The shape of the IRF has only a very minor effect on the results presented in this study. The main effect on the positional accuracies comes from the image ``width" and the steepness of its slopes, and a Gaussian profile forms a good first approximation for a typical beam profile.

\subsection{Samples and sample noise}
Two distributions of samples have been considered, one following a realistic scan-velocity, which will generally create an uneven distribution of samples over an image, and a ``smooth" sampling, with all samples spread evenly over the image. The latter was finally preferred for testing, as it should provide an simple-to-interpret set of results, not unnecessarily complicating the possible detection of systematic errors that may possibly be inherent to the methods tested. Thus, at each pointing direction a set of evenly spaced samples was created that represented the 60 scans made at that pointing. Gaussian noise was added to these samples at a level according to a specified signal to noise ratio, which applied to the peak of the image. The contributing scans for each source were determined by the adopted scanning law, which defines the orientation and position of each scan relative to that of the source. For each set of contributing transits of a given source, 100 different noise realizations were made.

\section{The processing of the scan data}
\label{sec:scandata}

The available simulated data for each source were investigated following the methods described in Section~\ref{sec:correst} through weighted least-squares solutions, providing estimated values for the position and flux (the image parameters), and the standard deviations on those values. These parameter values were compared with the input data, and the differences compared with the estimated standard errors. The weights used in the solutions were based on the standard deviation of the Gaussian noise used in the data simulations. The mean values of the image parameters and their standard errors were determined as based on the 100 independent noise realizations. This allows in principle for bias detection at 10 per~cent of the standard deviation. For a $S/N=10$, typical values of just under 2~arcsec were found for the positional standard error, with a sigma for the beam of 841~arcsec. For the flux, the standard error was approximately 0.15~per~cent. The observed dispersion in the measured values was in good agreement with the derived standard errors for the solutions. More details on results are provided in Section~\ref{sec:comparisons}.

\section{Image parameters from pixel maps}
\label{sec:pixmapdata}
As part of the experiment, the samples as obtained in the scans were projected on a map of square pixels. To avoid complications caused by distorted pixels, each source was projected as if it was situated at the ecliptic plane, where the map approaches a flat patch. The pixel size was chosen at 6 arcmin, which is compatible with the sample size on the scans (5.4~arcmin) and the maximum distance between successive scans (3.15~arcmin), and guarantees that at least two samples are contributing to each pixel. The 6~arcmin pixels correspond to what is referred to in the HEALPix\footnote{http://healpix.jpl.nasa.gov} software as $\mathrm{NSIDE}=2^9=512$ \citep{gorski05}. At the next resolution level, the pixel size would be 3~arcmin, leaving some pixels empty.

Two options were explored, one in which the source took a random position within a pixel, and a second where the source was central in a pixel. This allowed for some distinction of how pixelization may affect the extracted image parameters.

After defining the position of the source with respect to the map pixels, the samples from the accumulated scans were projected on the relevant pixels, where each sample was assigned in full to the nearest pixel, as is standard procedure for creating CMB pixel maps \citep[see for example][]{ashdown07}. At each pixel, the response was simply taken as the mean of all contributing samples. 

The image-parameter extraction for the pixel maps follows the same principles as for the scan data. We assumed in all cases a circular symmetric 2-dimensional Gaussian profile for the image, with the width defined by the mean $\sigma$ of the actual beam. In case of an asymmetric beam profile this will be somewhat inaccurate, but in that situation the image profile on the map is always poorly defined due to the different contributions to individual pixels. Thus, we start with a 2-dimensional Gaussian profile for the IRF as in Eq.~\ref{equ:2dgauss}, but now with respect to pixel coordinates $(x,y)$ and a first assumption of the source position $(x_0,y_0)$:
\begin{equation}
R(x,y) = \exp\biggl[-\frac{(x-x_0)^2}{2\sigma^2}-\frac{(y-y_0)^2}{2\sigma^2}\biggr]
\end{equation}
The predicted response for a source in pixel $(x,y)$ is simply given by:
\begin{equation}
\tilde{o}_{x,y} = \tilde{I}\cdot\exp\biggl[-\frac{(x-\tilde{x}_0)^2}{2\sigma^2}-\frac{(y-\tilde{y}_0)^2}{2\sigma^2}\biggr].
\end{equation}
Corrections to the estimated image parameters $(\tilde{x}_0, \tilde{y}_0, \tilde{I})$ are obtained in the same way as was done for the scan data, by comparing the predicted with the observed responses:
\begin{equation}
o_{x,y} - \tilde{o}_{x,y} = R(x,y)\:\mathrm{d}I + \tilde{I}\:\frac{\partial R}{\partial x}\mathrm{d}x + \tilde{I}\:\frac{\partial R}{\partial y}\mathrm{d}y.
\end{equation}
These equations are solved by least squares. Iterations are required as corrections for the position estimate affect the predicted responses in a non-linear way. The final iteration, based on a convergence criterion, provides a set of instrument parameters and their standard errors.

\section{Results and comparisons}
\label{sec:comparisons}

In the comparisons of results, the following performance criteria were looked at:
\begin{itemize}
\item Accuracy of the reproduction of the source position;
\item Accuracy of the predicted standard error on the source position;
\item Accuracy of the reproduction of the flux;
\item Accuracy of the predicted standard error on the flux.
\end{itemize}
Together these criteria define both the reliability and robustness of the methods that have been applied.
 \subsection{Distribution of test sources}
Tests were carried out with sources distributed evenly over all ecliptic longitudes at 1~degree separation and at a few selected fixed ecliptic latitudes. The scan coverage of a source is a strong function of both ecliptic latitude and longitude, and any algorithm applied to source-parameter extraction will need to be able to provide reliable information under all these conditions. This is particularly the case for a relatively short mission such as Planck, where the mission is expected to last less than 2 years. The local variations in the distribution of scan directions also become more pronounced for a Planck-like scan for which the solar aspect angle is small. 

\subsection{The test data sets}

The test data sets we created aimed at testing the methods under increasingly more complex and difficult conditions. The first data set, described in Section~\ref{sec:circbeam}, used a circular-symmetric beam and $S/N=10$. Map pixelization was done by randomly positioning of the source with respect to the pixels on the ``map''. Data were obtained at ecliptic latitudes of 0 and 60 degrees.

In Section~\ref{sec:circ_fixed} the pixel positions were forced such that the source would be positioned exactly at the centre of a pixel. Here data were only obtained a latitude of 0 degrees.

Finally in Section~\ref{sec:var_ellip} the most extensive tests are described. These covered 4 latitude (0, 30, 60 and 89 degrees), 4 ellipticities (1.0, 1.2, 1.4 and 1.6) and 7 signal-to-noise ratios ($S/N=0.1, 0.2, 0.5, 1.0, 3.0, 5.0, 10.0$), equivalent to an intensity range from 0.01 to 100. There was only one longitude used, $\lambda=240$~degrees, which had shown in circular-beam tests to be optimal for the map-making results, independent of pixel positions.

\subsection{Circular-symmetric beam}
\label{sec:circbeam}
This data set provides what should be the easiest conditions for the pixel-map extractions, and can show the minimum difference in performance between the two methods. 
\begin{figure}
\centering
\includegraphics[width=8.3cm]{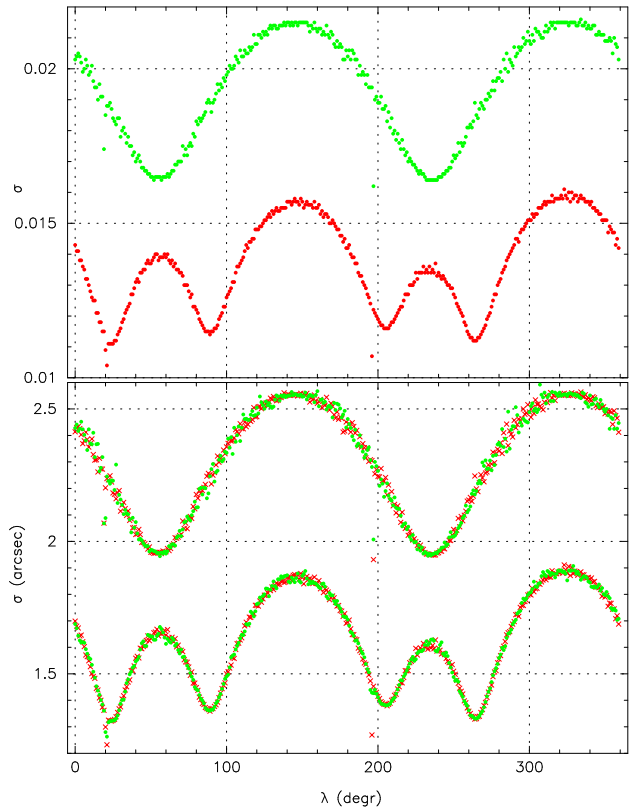}
\caption{Standard errors for scan-based solutions as a function of ecliptic longitude, in both graphs for latitudes of 0 (top) and 60 (bottom) degrees. The lower graph shows the results in longitude (crosses) and latitude (dots), which  are observed to be very similar. This could be expected for a circular symmetric beam. The few points showing in each graph higher than average accuracies are due to an overlap at the closure of the full scan. The top graph shows the same for the flux determinations, where the errors need to be compared with a peak flux of 10.0.}
\label{fig:ringsterr}
\end{figure}
\subsubsection{The scan-based analysis}
All results for the scan-data analysis are internally consistent and at, or very close to, the statistical limit. The mean standard errors over the 100 noise realizations as a function of ecliptic longitude display the variations in density of scan coverage (Fig.~\ref{fig:ringsterr}). One of these variations is caused by the smaller than average displacements between scans when the spin-axis movements are opposite to the apparent movement of the Sun (causing the denser scan at around 55 and 235 degrees). As was stated already in Section~\ref{sec:scanlaw}, when those movements are in the same direction, the scan is less dense and the standard errors are larger (at longitudes of 145 and 325 degrees). Also visible are a couple of data points affected by the ``full closure" of the scan, causing effectively a few additional scans covering longitudes at 20 and 200 degrees. Very similar curves are observed for the standard errors on the flux, as is shown in the top graph of Fig.~\ref{fig:ringsterr}. A $\sigma$ value of 0.02 is equivalent to a 0.2~per~cent error on the flux measurement.
\begin{figure}
\centering
\includegraphics[width=8cm]{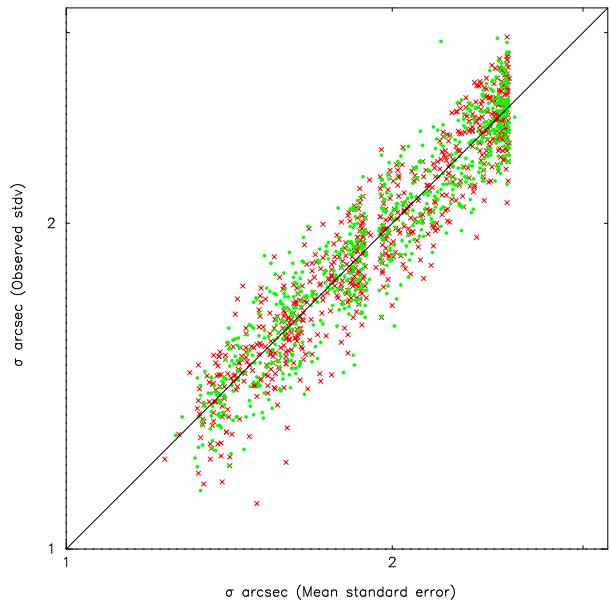}
\caption{Comparison between the mean standard errors and observed standard deviations for 100 noise realisations for the scan-based solutions. The data in longitude (crosses) and latitude (dots) are shown for both the latitude 0 and 60 simulations.}
\label{fig:ringsdvcomp}
\end{figure}
The next test is to see how the mean standard errors for each longitude compare with the observed standard deviations, as based on the analysis of the 100 noise realizations (Fig.~\ref{fig:ringsdvcomp}). The spread of the data points is close to what could be expected for standard deviations based on 100 independent measurements. We observe a small increase of the standard deviations for the lowest fluxes. The same is observed for the errors on the flux measurements. There the increase accompanies the development of a positive bias in the flux estimates. 

\begin{table}
\caption{Reproduction of input data with scan-based solutions}
\begin{tabular}{|l|rrrr|}
\hline
Coord. & Resid. & uwsd & Resid. & uwsd \\
 & \multicolumn{2}{c}{$\beta=0$} &  \multicolumn{2}{c}{$\beta=60$} \\
\hline
$\delta\lambda\cos\beta$ & $0.003\pm0.012$ & 1.024 & $0.011\pm0.008$ & 1.009 \\
$\delta\beta$ & $0.012\pm0.012$ & 0.932 & $-0.010\pm0.008$ & 1.003 \\
$\frac{I-10}{0.01}$ & $-0.003\pm0.010$ & 0.943 & $-0.009\pm0.007$ & 0.972 \\
\hline
\end{tabular}
\begin{list}{}{}
\item uwsd:unit-weight standard deviation
\end{list}
\label{tab:ringresid}
\end{table}
The final comparison is between the input and observed values for the positions and fluxes. Values have been obtained while averaging over all longitudes. Small and insignificant differences are seen depending on whether the observed standard deviations or the mean standard errors are used as weights. Given the close agreement between these quantities, this is no surprise. The errors on the mean values per longitude have been taken as 0.1 times the mean standard error, considering that 100 observations were used to calculate each of the means. The results are summarized in Table~\ref{tab:ringresid}, which shows that there are no biases observed in the scan-based analysis results for either positions or fluxes.

\begin{figure}
\centering
\includegraphics[width=8cm]{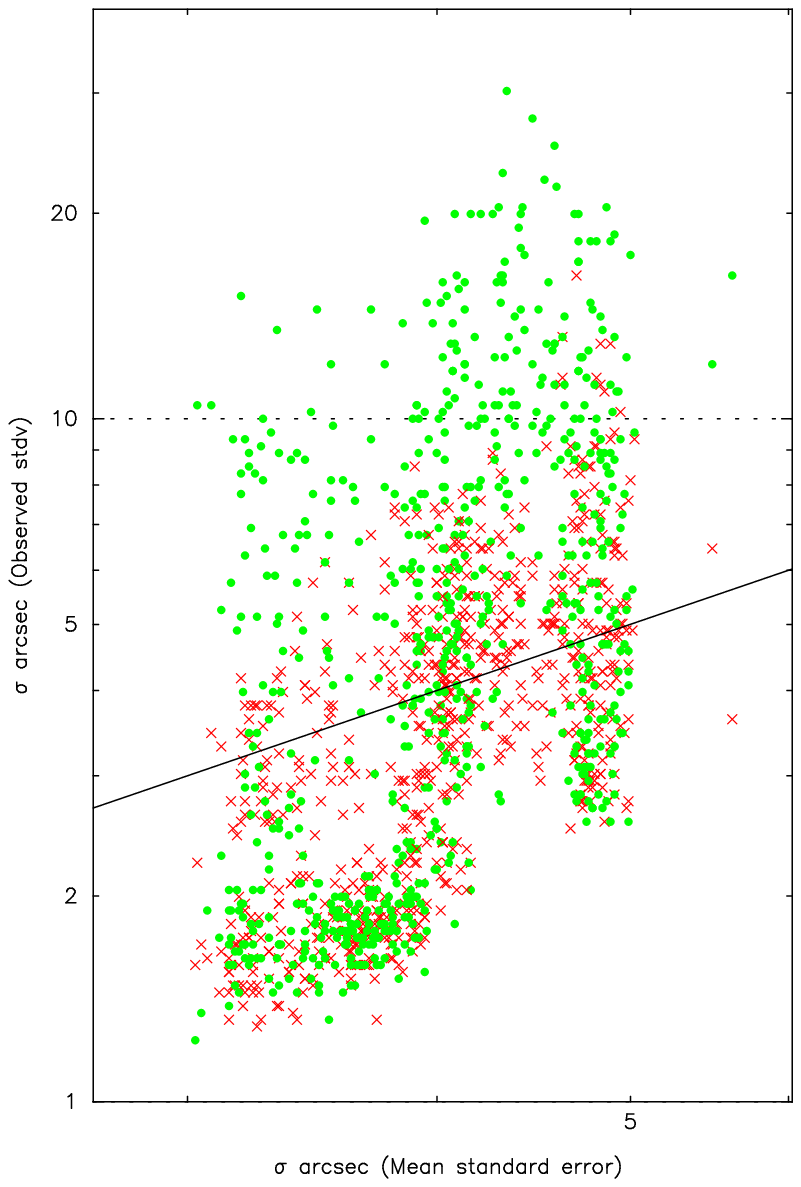}
\caption{Comparison between the mean standard errors and observed standard deviations for the pixel-map based solutions. The diagonal line shows the one-to-one relation. Dots and crosses refer to measurements in latitude and longitude respectively.}
\label{fig:mapsdvcomp}
\end{figure}
\subsubsection{The pixel-map based analysis}
The results for the pixel-map based analysis are less accurate or predictable. We observe a poor relation between the standard errors of the solutions and the actual observed standard deviations, as shown in Fig.~\ref{fig:mapsdvcomp}. The comparison shows that the standard errors derived from the least-squares solutions generally provide in this situation a very poor estimate of the actual errors on the measurements. A comparison with Fig.~\ref{fig:ringsdvcomp} shows that for the analysis of the pixel-map data the positional errors are mostly larger, by up to a factor ten, than the scan-based analysis of the same data.
\begin{figure*}
\centering
\includegraphics[width=14cm]{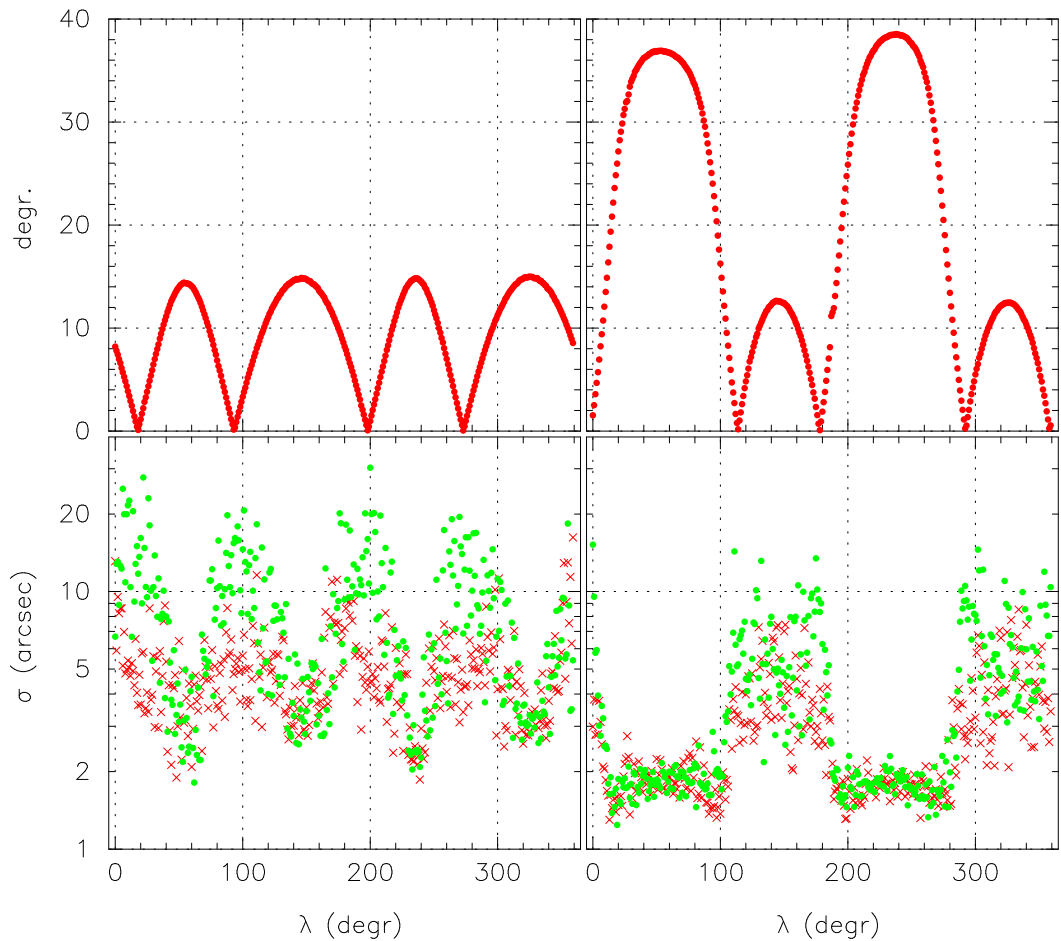}
\caption{Lower graphs: the observed standard deviations for pixel-map based solutions at latitude 0 (left) and latitude 60 (right). Upper graphs: the scan intercept angle (see text) as a function of ecliptic longitude. Dots and crosses refer to measurements in latitude and longitude respectively.}
\label{fig:mapsdvlam}
\end{figure*}
\begin{table}
\caption{Reproduction of input data with pixel-map based solutions}
\begin{tabular}{|l|rrrr|}
\hline
Coord. & Resid. & uwsd & Resid. & uwsd \\
 & \multicolumn{2}{c}{$\beta=0$} &  \multicolumn{2}{c}{$\beta=60$} \\
\hline
$\delta\lambda\cos\beta$ & $-0.019\pm0.021$ & 0.957 & $-0.115\pm0.011$ & 1.361 \\
$\delta\beta$ & $0.002\pm0.025$ & 1.412 & $0.016\pm0.012$ & 2.393 \\
$\frac{I-10}{0.01}$ & $-7.4\pm0.18$ & 0.943 & $-7.4\pm0.093$ & 2.298 \\
\hline
\end{tabular}
\begin{list}{}{}
\item uwsd:unit-weight standard deviation
\end{list}
\label{tab:madresid}
\end{table}
The main problems with the analysis of the pixel-map data are caused by poor scan coverage. Each source on the ecliptic plane will generally receive two scans over a 12 months period. At the ecliptic plane these scans will have a difference in inclination by between 0 and 15 degrees (twice the solar aspect angle of the spin axis). This is referred to as the intercept angle. For values close to zero, the standard deviation for the solution tends to be large, while at maximum value the pixel-map based results become close to those of the scan-based analysis (Fig.~\ref{fig:mapsdvlam}). The same effects are observed in the errors for the flux measurements. However, while there is no bias in the positional measurements observed, there is a significant flux loss (see Table~\ref{tab:madresid}). A simple comparison between the data in Tables~\ref{tab:ringresid} and \ref{tab:madresid} as well as between Figures~\ref{fig:ringsdvcomp} and \ref{fig:mapsdvcomp} clearly demonstrates the performance advantages of the scan-based analysis, providing reliable and generally smaller standard errors and a bias-free flux estimate, with no noticeable dependence on the intercept angle (strictly speaking, an intercept angle can only be defined when two sequences of related scans are present).

\subsection{Circular symmetric beam, source position fixed at pixel centre}
\label{sec:circ_fixed}
\begin{figure}
\centering
\includegraphics[width=8.3cm]{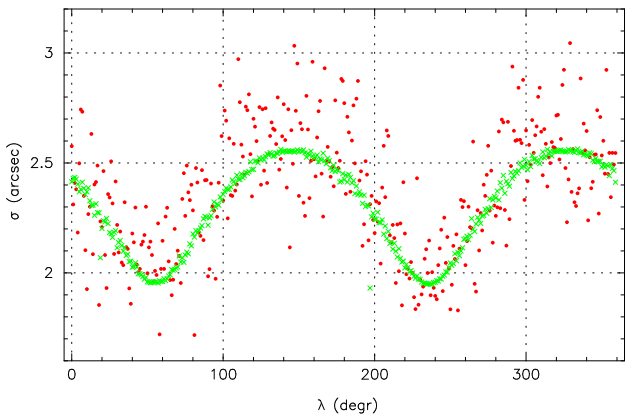}
\caption{Positional standard deviations over 100 noise realisations, as a function of ecliptic longitude, for sources at ecliptic latitude zero. The broad distribution of dots are the results for pixel-map based positions, where the source was always forced to be at the centre of a pixel. The narrow distribution of crosses shows the same data analysed directly from the scans.}
\label{fig:mapfxlam}
\end{figure}
\begin{figure}
\centering
\includegraphics[width=7cm]{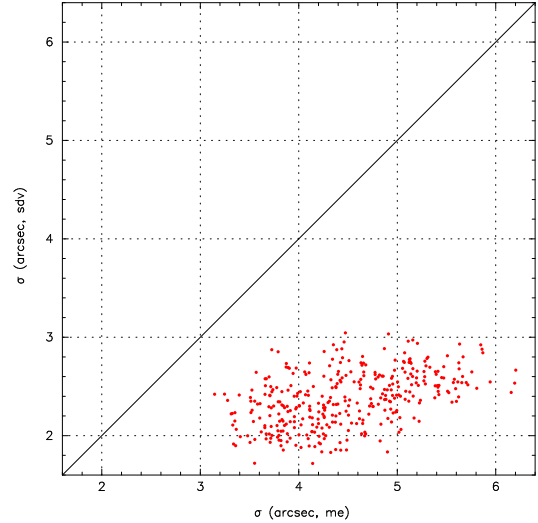}
\caption{Observed positional standard deviations compared with mean standard errors from the image-parameter solutions.}
\label{fig:mapfxsdv}
\end{figure}
\begin{figure}
\centering
\includegraphics[width=8.3cm]{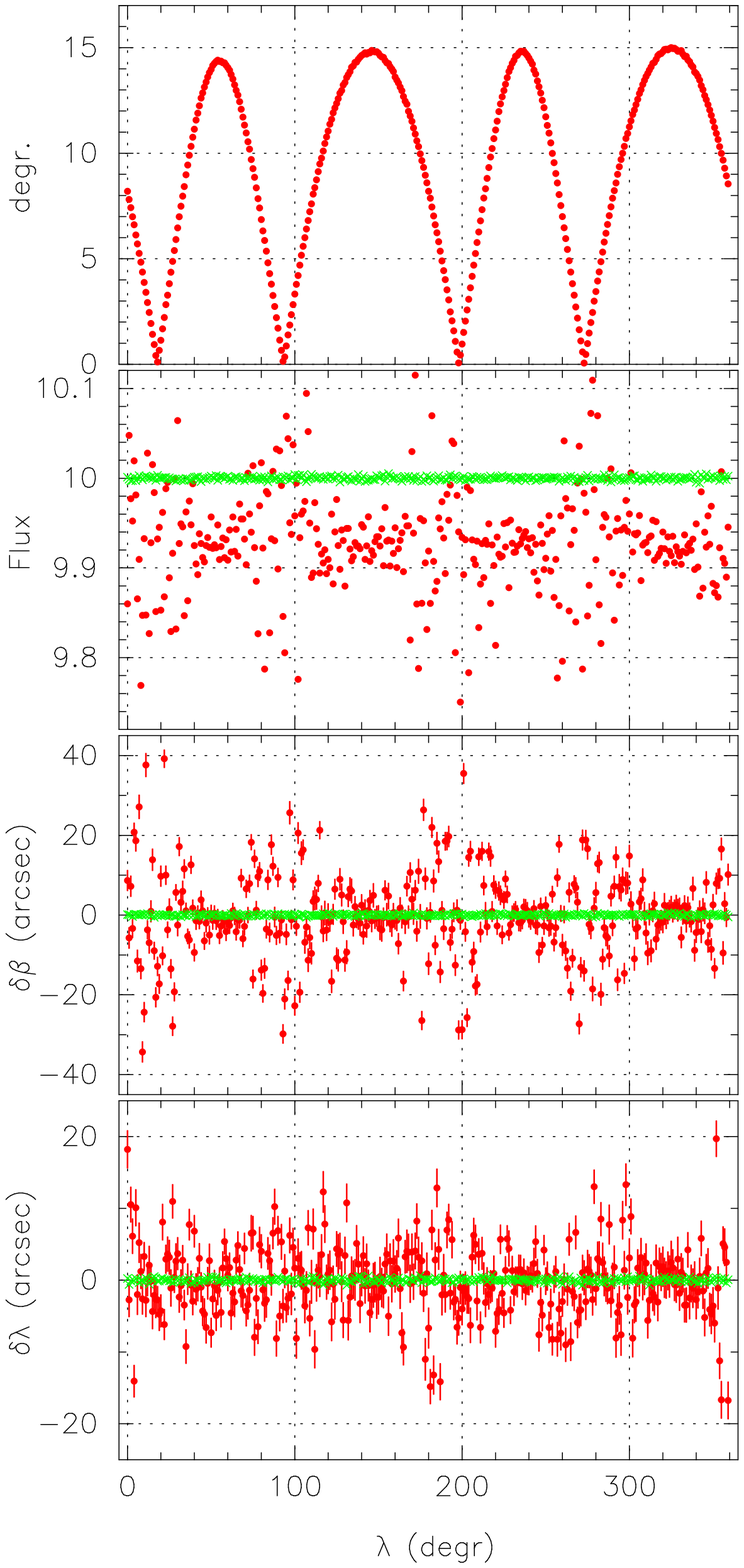}
\caption{Lower three graphs, from top to bottom: flux measurements, residuals in latitude and residuals in longitude. The wide spread of points in each graph originate from the pixel-map based analysis, the narrow distributions from the scan-based analysis. Upper graph: the intercept angle of the scans contributing to the data. All data are shown as a function of ecliptic longitude and concerns sources at ecliptic latitude zero.}
\label{fig:mapfimeans}
\end{figure}
\begin{figure}
\centering
\includegraphics[width=7.8cm]{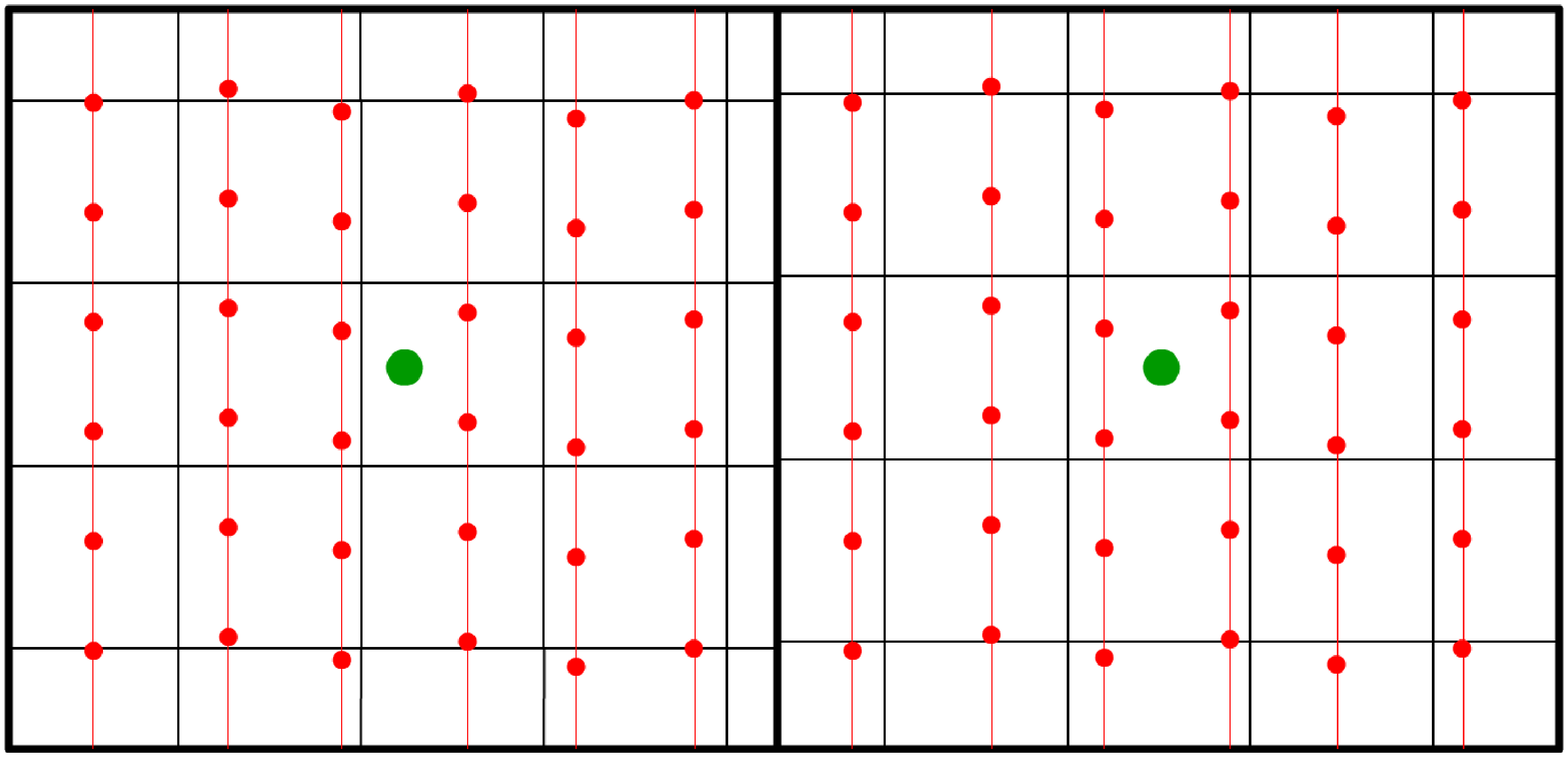}
\caption{Examples of a pixel grid coverage of scan observations, showing the coverage variations due to accidental positioning of the grid with respect to the source position. The central dot represents the source position. The connected sequences of small dots represent samples on successive scans.}
\label{fig:pixelgrid}
\end{figure}

In order to investigate the origin of the rather large standard deviations observed in the map-based solutions, a set of solutions at latitude 0 degrees was made such that the image was always positioned at the centre of a pixel on the map. The standard deviations of the positional fits are now closer to what is found for the scan data (Fig.~\ref{fig:mapfxlam}), though the noise level on the map data is on average higher and repeatability of those data is a lot poorer. The relation between those standard deviations and the standard errors is in addition still very poor too. The mean standard error estimates following from the least squares solutions of the map data are systematically overestimated and generally very noisy (Fig.~\ref{fig:mapfxsdv}). Thus, even in these, what may appear to be relatively favourable, but still unrealistic, conditions, the pixel-map derived positional information is still poor, in the prediction of the errors, while the errors themselves are poor estimates of the actual noise on the data. This can be observed in Fig.~\ref{fig:mapfimeans}, where the offsets for the recovered positions in longitude and latitude are shown as a function of longitude of the source. For comparison, the results for the scan-based analysis are also shown. An example of how the distribution of pixels affects the data accumulated from the scans is shown in Fig.~\ref{fig:pixelgrid}. A summary of these results as well as a comparison with the scan-based analysis is presented in Table~\ref{tab:mapfresid}. Here both the actual standard deviation and the unit-weight standard deviation are shown for the same data analysed from pixel maps and directly from scan data. The conclusion can be drawn that the scan-based analysis is performing more than an order of magnitude better than the pixel-map based analysis, in the reconstruction of all image parameters as well as in the provision of standard errors. 
\begin{table}
\caption{Reproduction of input data with pixel-map based solutions for fixed source positions}
\begin{tabular}{|l|rrrr|l}
\hline
Coord. & uwsd & sd & uwsd & sd & Units\\
 & \multicolumn{2}{c}{Pixel-map} &  \multicolumn{2}{c}{Scans} \\
\hline
$\delta\lambda\cos\beta$ & 1.106 & 4.92 & 1.024 & 0.237 & arcsec\\
$\delta\beta$ & 2.518 & 10.43 & 0.932 & 0.213 & arcsec\\
$\frac{I-10}{0.01}$ & 1.547 & 5.50 & 0.972  & 0.185 & \\
\hline
\end{tabular}
\label{tab:mapfresid}
\end{table}

Also in flux reconstruction the results for the pixel-map derived solutions remain poor, as is shown in Fig.~\ref{fig:mapfimeans}. There is still a systematic flux loss of on average about 7~per~cent, and a noise that is in particular high for small intercept angles. None of these features shows in the analysis directly made from the scan data, the results of which are also shown in Fig.~\ref{fig:mapfimeans}.
Conclusions we can draw from these experiments are that the position of the source centre with respect to the pixels on the pixel map adds a significant source of noise, in particular for small intercept angles. But even when this ``positional freedom" is removed, and the sources are artificially placed at pixel centres, the results remain significantly inferior to those obtained directly from the scan data, in reproduction of the image parameters (with much higher noise and a flux loss for the map-based analysis) and in accuracy of the standard error estimates. Considering that we are still dealing here with a circular-symmetric beam, it appears that the actual pixelisation process is a major source of additional noise in the source-parameter determination.  

\subsection{Varying ellipticity, flux and latitude}
\label{sec:var_ellip}
\begin{figure*}
\centering
\includegraphics[width=15cm]{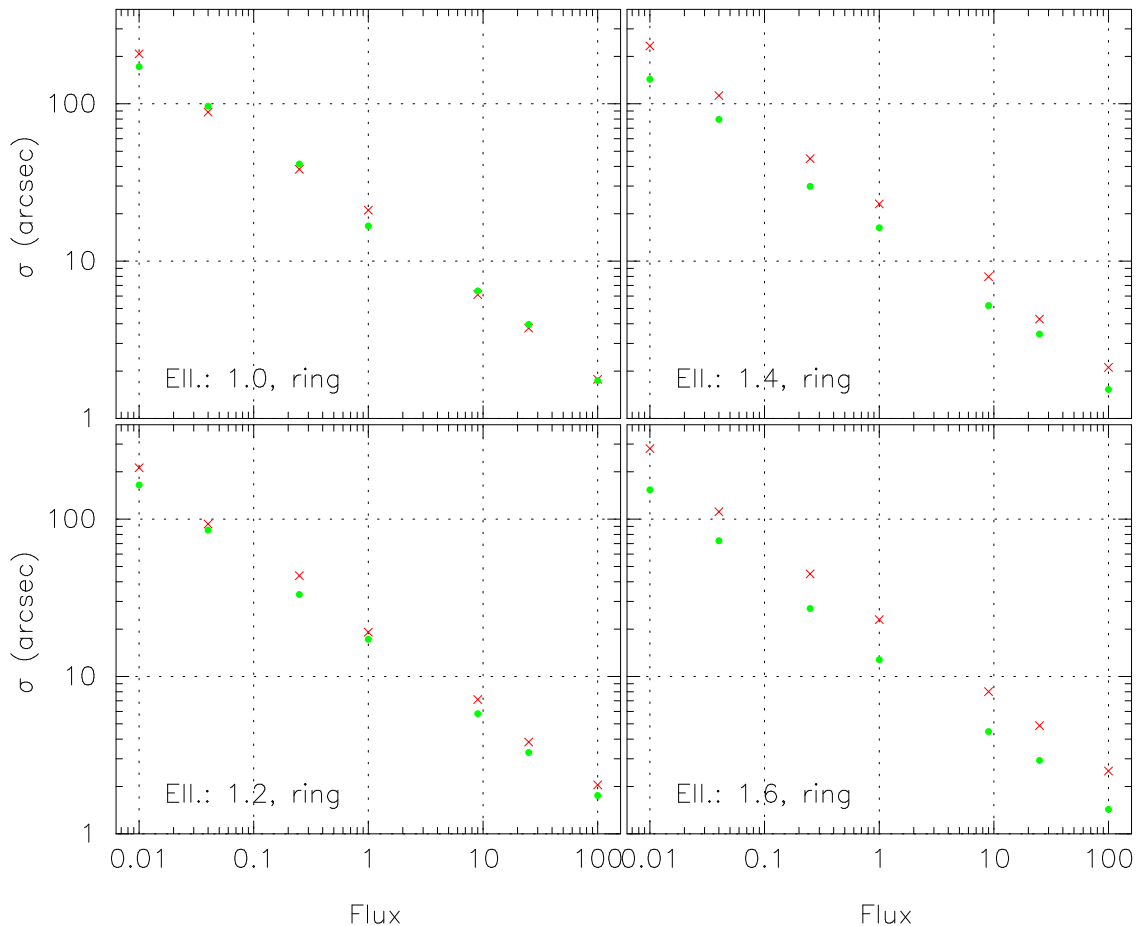}
\caption{Standard deviations of positions for different ellipticity, based on the direct analysis of the scan data, as a function of source intensity. Data in longitude direction are shown as crosses, in latitude direction as dots. The differences between the longitude and latitude results for higher ellipticity values reflect the average beam shape. In this example the source was positioned at 30 degrees latitude.}
\label{fig:ellring30}
\end{figure*}
For the final test runs, on the effects of ellipticity on the image-parameter reconstruction, an ecliptic longitude of 240~degrees on the sky was chosen for which the pixel-map based analysis was giving the relatively best results. Sources were placed at latitudes of zero, 30, 60 and 89~degrees. The fluxes of sources at those positions were varied over a range of S/N values from 0.1 to 10, equivalent to a range of source flux from 0.01 to 100. The ellipticity, defined as the ratio of the FWHM along scan over the FWHM across scan, was tested for values of 1.0, 1.2, 1.4 and 1.6, where a value of 1.4 is typical for the lowest frequency detectors on the Planck satellite \citep{sandri05}. Each point in this 3-dimensional data grid was simulated with 100 independent noise realizations, The simulated data were analysed from the scans and projected onto pixel maps.

\begin{figure}
\centering
\includegraphics[width=7cm]{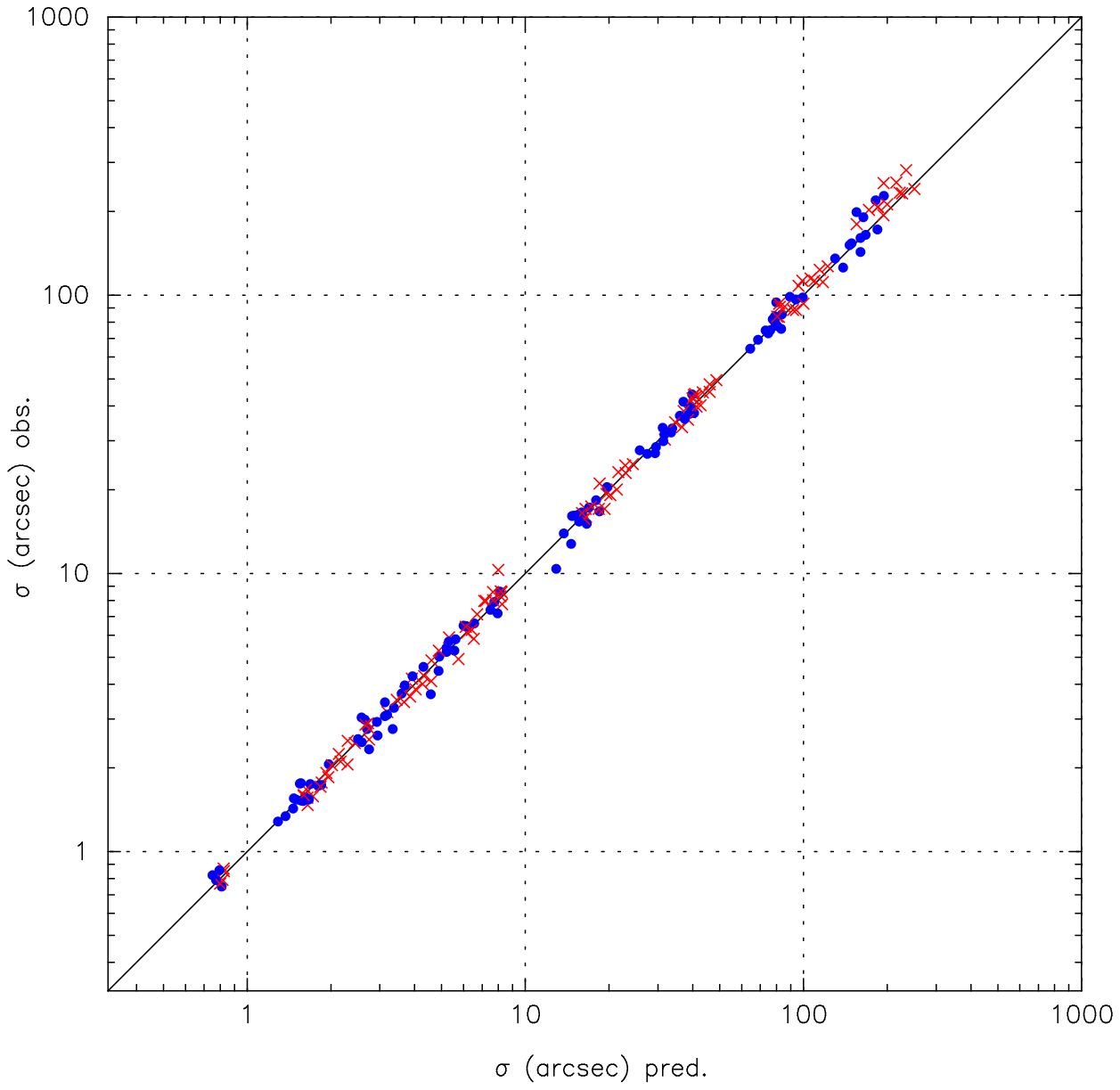}
\caption{A comparison between measured standard deviations and mean standard errors for the scan-based solutions for all test cases. Only for the faintest sources (at a S/N on the scan of 0.01) is there a small increase in the observed with respect to the predicted noise level. The crosses and dots refer to the data for longitude and latitude respectively.}
\label{fig:sdvringell}
\end{figure}
Figure~\ref{fig:ellring30} shows an example of the results from the scan-based analysis, in this case for a source at latitude 30 degrees. Results from the scan-based analysis for other latitudes were very similar and mainly reflected difference in total scan coverage. As can be observed from Fig.~\ref{fig:ellring30}, the standard deviations of the fitted positions as obtained from the different noise realizations behaves fully in the expected way, i.e.\ they are proportional to the square root of the flux. The only additional feature visible is a split between the standard deviations for the longitude and latitude directions for the high-ellipticity beam profiles, which simply reflects the preferential direction of scans being close to perpendicular to the ecliptic plane. The effective beam width in longitude is therefore narrower than in latitude.

\begin{figure}
\centering
\includegraphics[width=7.5cm]{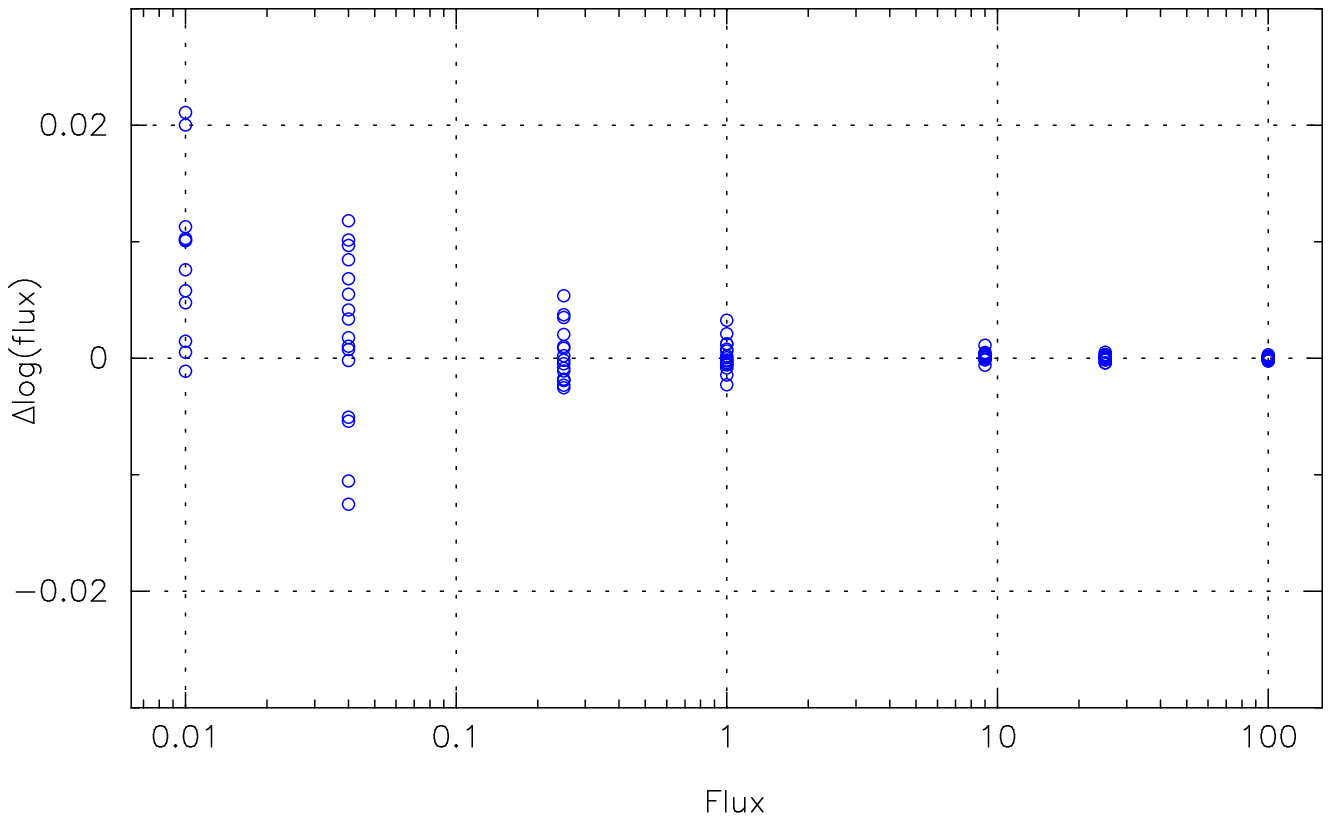}
\caption{Recovery of the input flux in the scan-based solution as a function of the source flux on a single scan. All 16 test cases (4 latitudes times 4 ellipticity values) are shown for each flux. Multiplication of the vertical scale by 2.5 gives the equivalent errors in magnitudes.}
\label{fig:fluxring}
\end{figure}
Equally important for the analysis is the reliability of the standard errors. A comparison between the standard errors and the observed standard deviations for all test cases as obtained from the scan-based analysis is shown in Fig.~\ref{fig:sdvringell}. No dependencies on latitude or ellipticity are observed, and there is only a small increase of the observed standard deviations compared to the mean standard errors for the faintest sources.

The errors on the reconstructed flux values are shown in Fig.~\ref{fig:fluxring}.  A small bias is observed for the lowest two flux levels, which doesn't show a dependence on either latitude of the source or ellipticity of the beam. No significant bias is observed for sources at flux level 1 and above, i.e.\ those with S/N$\ge 1$ on the original scan data. The errors on the fluxes are equivalent to an accuracy of order 0.01~magn.\ or better for all sources with a S/N$\ge 1$ in the scan data.

Exactly the same data has been analysed after projection on a pixel map as described earlier in this paper. No special provisions were made about placement of the source position with respect to the pixels, but the longitude of the sources had been chosen such that what appeared to be the most favourable conditions for the pixel-map based solutions would be obtained. At almost any other longitude the tests described in Section~\ref{sec:circbeam} showed results that are (considerably) worse. In contrast, a similar variation in longitude was shown to have only very small effects (only due to total coverage variations) on the direct analysis of the scan data. 
\begin{figure*}
\centering
\includegraphics[width=15cm]{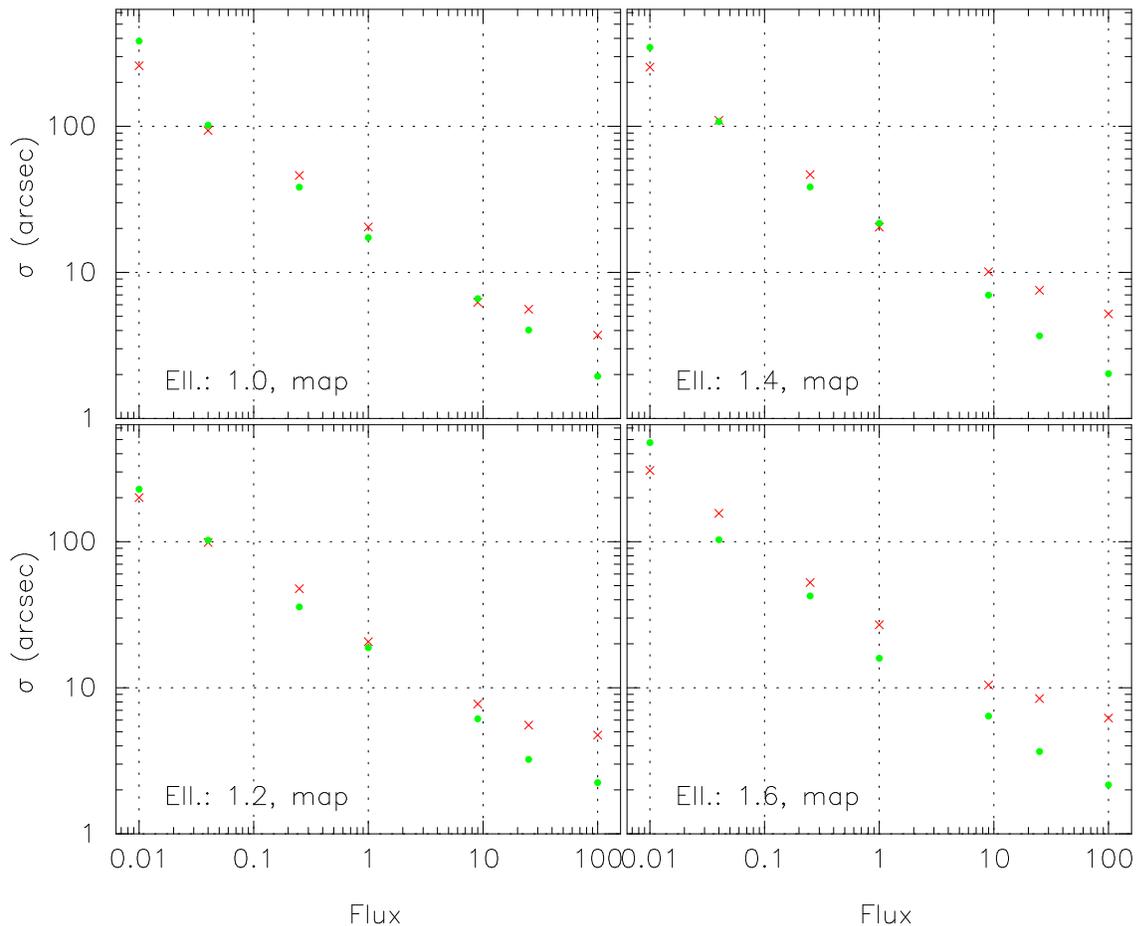}
\caption{The same data as shown in Fig.~\ref{fig:ellring30}, but now analysed from the pixel map. }
\label{fig:ellmap30}
\end{figure*}

Figure~\ref{fig:ellmap30} shows the equivalent of Fig.~\ref{fig:ellring30}, but now for the pixel-map based analysis. A couple of differences can be noticed. For the brightest images the sensitivity to the ellipticity of the beam has an increasingly damaging effect, which is what could be expected, as the image-parameter fits for brighter images become increasingly more sensitive to an accurate beam representation. For all sources the errors are somewhat larger than for the scan-based analysis, but this is in particular the case for the fainter ones. It should be noted too that in the pixel-map based analysis between 4 and 8~per~cent of the parameter fits on the faintest images were unsuccessful (i.e.\ didn't converge). No such failures were found in the scan-based analysis. 

\begin{figure}
\centering
\includegraphics[width=7cm]{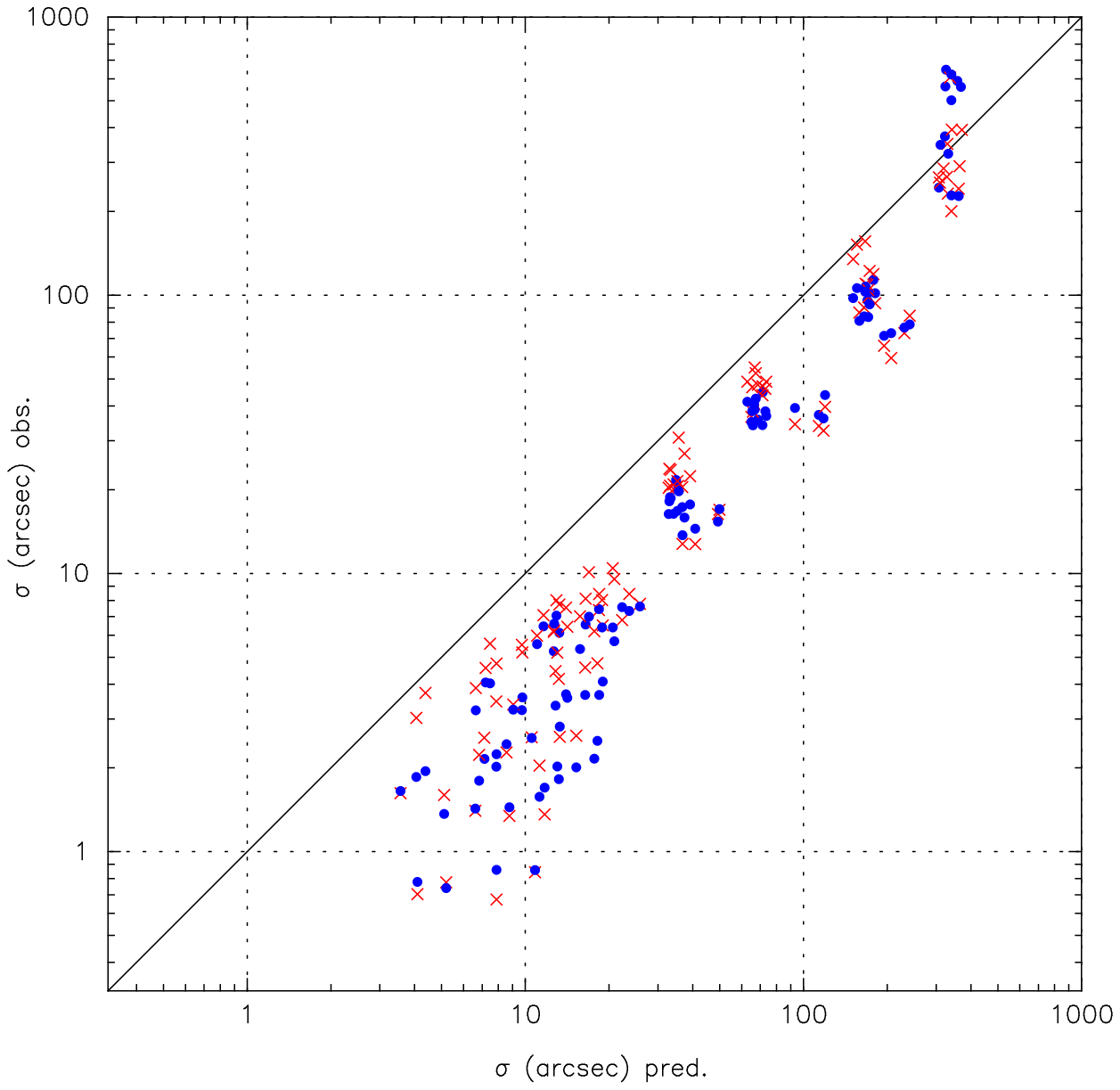}
\caption{A comparison between mean standard errors and observed standard deviations for the pixel-map based analysis. Data are shown for all test cases. The equivalent data for the scan-based analysis is shown in Fig.~\ref{fig:sdvringell}. The crosses and dots refer to the data for longitude and latitude respectively.}
\label{fig:sdvmapell}
\end{figure}
As was observed for the circular beam analysis, the standard errors obtained from the pixel-map based solutions are inaccurate, and tend to be, at this longitude, an overestimate. This is seen here in Fig.~\ref{fig:sdvmapell}, which, however, also shows that for the faintest images the actual standard deviations can be larger than the mean standard errors obtained from the solutions, similar to what is observed for the scan-based analysis.  

\begin{figure}
\centering
\includegraphics[width=7.5cm]{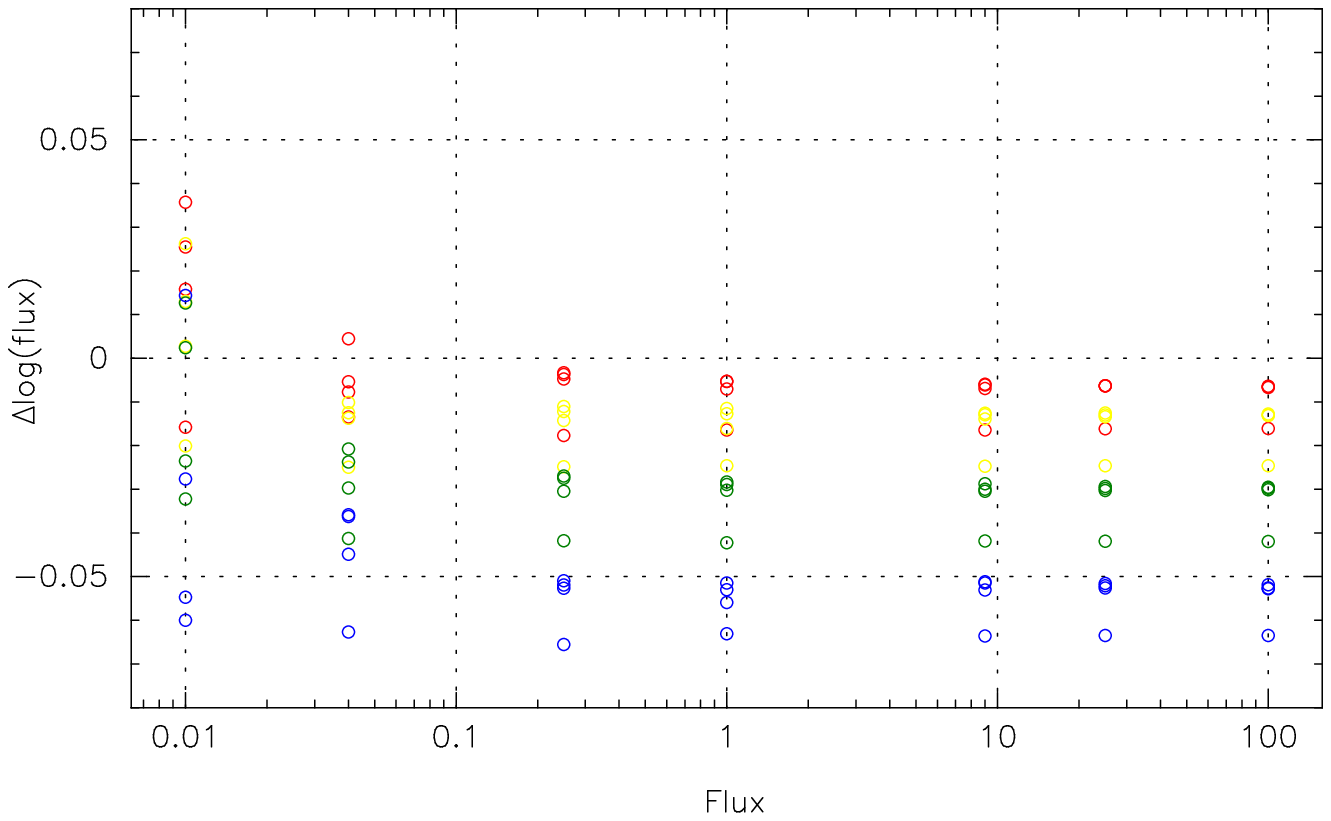}
\caption{The equivalent of Fig.~\ref{fig:fluxring}, but now for the pixel-map based analysis. The spread in data points for each flux value is closely related to the ellipticity (indicated by colour, from low, red, to high, blue), with a secondary dependence on latitude, such that the worst bias is, as could be expected, found for the highest ellipticity, while sources at higher latitude are less affected than those at lower latitudes.}
\label{fig:fluxmap}
\end{figure}
Figure~\ref{fig:fluxmap} shows the reconstruction of the flux values for the pixel-map based analysis, and should be compared with Fig.~\ref{fig:fluxring} of the scan-based analysis. As was observed in Section~\ref{sec:circbeam}, the reconstructed fluxes in the pixel-map based analysis are biased, systematically providing too low estimates of the actual flux. This becomes worse in the case of an elliptical beam, and in particular close to the ecliptic plane, where the image on the map is constructed from only scans at a narrow range of intercept angles. In magnitude terms, the bias would be of order +0.1 for an ellipticity of 1.4 to 1.6.

In conclusion, the pixel-map based analysis for even the most favourable map data is still significantly less accurate than the scan-based analysis for any of the data. This applies to all aspects of the image-parameter extraction: positions, fluxes and the standard errors on the estimated parameters.  

\section{Discussion and conclusions}
\label{sec:discussion}
Next to the frequency maps, the compact source catalogue provides a lasting heritage of a survey mission such as Planck. It should therefore be prepared with the utmost care and to the best capacity of the actual data, so that it can be reliably compared with survey data and source catalogues in other frequency domains. 

It appears that the main source of noise in the pixel-map based extraction of image parameters originates from the manner in which the data are projected on the pixel map. However, any more sophisticated method, considering for example a spread of data over several pixels, will be complicated and confuse the statistical properties of the data.

We have shown that there is a significant advantage in extracting source parameters from a full-sky scanning survey mission directly from the scans rather than from the data as projected on a pixel map. The methods themselves are simple to implement, are stable, and provide accurate results for positional and flux information on the sources as well as reliable standard errors on these parameters. The methods as described here are easily extendible to multiple frequencies and can also take into account background gradients. These are trivial extensions of the implementation we have tested here. As a further extension, but somewhat more complicated and requiring further tests, an implementation could also include a final adjustment of what is generally referred to as the instrument parameters, describing the relative positions and flux-responses of a set of different detectors in the focal plane. 

Extending the analysis of the image parameters across different frequency maps has the added advantage of producing source information that is always consistent across different maps, and could even take fully into account situations where two sources are resolved on a high-resolution map, and appear merged on a low resolution map, providing well-defined separate fluxes on all maps.

A further advantage of the scan-based over the map-based analysis is the ability of the former to determine the source flux over a range of epochs, determined by the frequency the source is transitted rather than the frequency at which the maps may be made.

A good strategy for the creation of a complete and reliable source catalogue for a survey mission could be the following:
\begin{enumerate}
\item After a proper geometric calibration of the instrument, prepare the frequency maps from the scan data; 
\item In the maps, at all frequencies, identify any feature that might be due to a compact source. This catalogue may be overcomplete, in other words may contain spurious detections, since these will be resolved and removed by the scan-based analysis;
\item Extract a list of potential source positions for all maps together;
\item Extract for each potential source the scan data at each frequency that would be within the beam-range of the source;
\item Apply the scan-based solution on the accumulated data for all frequency maps together, resolving any merged images where possible on the basis of higher resolution data;
\item Optionally, examine residuals for individual detectors on systematics that can reflect small corrections to the geometric and response calibrations; if these are found to be significant, apply and rebuild the pixel maps, then repeat points (ii) to (v).   
\end{enumerate} 
What will be obtained this way is a compact source catalogue that is reliable in positional and flux data, as well as the errors assigned to those data, and which is in addition fully consistent over all frequencies and very likely to be as complete as it can possibly be within the detection limits. In other words, a compact source catalogue that can have lasting value. It is also a catalogue that will automatically be free from any ambiguities between the results for different frequency maps.

\section*{Acknowledgments} Some of the results in this paper have been derived using the HEALPix \citep{gorski05} package. ANM is grateful for the support provided by a Marshall Scholarship and an NSF Graduate Research Fellowship during the preparation of this work

\bibliographystyle{mn2e}
\bibliography{MyBibliogr}

%\appendix
%
%\section[]{Large gaps in L\lowercase{y}${\balpha}$ forests\\* due to fluctuations in line distribution}

\label{lastpage}

\end{document}